\documentstyle[amsfonts,12pt,twoside,graphicx]{article}

\normalbaselines
\font\tbf = cmbx12

\begin{document}

\title{RELATIVISTIC EPICYCLES: \\
another approach to geodesic deviations }
\author{\tbf R. Kerner$^1$\thanks{
e-mail: {\tt rk@ccr.jussieu.fr}}, J.W. van Holten$^{2}$\thanks{
e-mail: {\tt v.holten@nikhef.nl}} and R. Colistete Jr.$^1$\thanks{
e-mail: {\tt coliste@ccr.jussieu.fr}} \\
\\
\mbox{\small $^1$Univ. Pierre et Marie Curie -- CNRS ESA 7065}\\
\mbox{\small Laboratoire de Gravitation et Cosmologie Relativistes}\\
\mbox{\small (as of jan.\ 1, 2001: L.P.T.L., CNRS URA 7600)}\\
\mbox{\small Tour 22, 4\`eme \'etage, Bo\^{\i}te 142, 4 place Jussieu,
      75005 Paris, France}\\
\mbox{\small $^2$Theoretical Physics Group, NIKHEF}\\
\mbox{\small P.O.Box 41882, 1009 DB Amsterdam, the Netherlands.}}
\date{\today}
\maketitle

\begin{abstract}
{\small We solve the geodesic deviation equations for the orbital motions in
the Schwarzschild metric which are close to a circular orbit. It turns out
that in this particular case the equations reduce to a linear system, which
after diagonalization describes just a collection of harmonic oscillators,
with two characteristic frequencies. The new geodesic obtained by adding
this solution to the circular one, describes not only the linear
approximation of Kepler's laws, but gives also the right value of the
perihelion advance (in the limit of almost circular orbits). We derive also
the equations for higher-order deviations and show how these equations lead
to better approximations, including the non-linear effects. The approximate
orbital solutions are then inserted into the quadrupole formula to estimate
the gravitational radiation from non-circular orbits. }
\end{abstract}

\vspace{0.7cm} PACS number(s): 04.25.--g, 04.30.Db

\newpage

\section{Introduction}

The problem of motion of planets in General Relativity, considered as test
particles moving along geodesic lines in the metric of Schwarzschild's
solution, has been solved in an approximate way by Einstein \cite{AE}, who
found that the perihelion advance during one revolution is given in the
near-Keplerian limit by the formula 
\begin{equation}
\Delta \phi =\frac{6\pi GM}{a(1-e^{2})}  \label{advance1}
\end{equation}
where $G$ is Newton's gravitational constant, $M$ the mass of the central
body, $a$ the greater half-axis of planet's orbit and $e$ its eccentricity. 
\newline
\indent
This formula is deduced from the exact solution of the General Relativistic
problem of motion of a test particle in the field of Schwarzschild metric,
which leads to the expression of the angular variable $\varphi $ as an
elliptic integral, which is then evaluated after expansion of the integrand
in terms of powers of the small quantity $\frac{GM}{r}$. \newline
\indent
The formula has been successfully confronted with observation, giving
excellent fits not only for the orbits with small eccentricities (e.g., one
of the highest values of $e$ displayed by the orbit of Mercury, is $e=0.2056$
), but also in the case when $e$ is very high, as for the asteroid Icarus ($%
e=0.827$), and represents one of the best confirmations of Einstein's theory
of gravitation. In the case of small eccentricities the formula (\ref
{advance1}) can be developed into a power series: 
\begin{equation}
\Delta \phi =\frac{6\pi GM}{a}\,(1+e^{2}+e^{4}+e^{6}+\dots ).
\label{advance2}
\end{equation}
One can note at this point that even for the case of planet Mercury, the
series truncated at the second term, i.e., taking into account only the
factor $(1+e^{2})$ will lead to the result that differs only by $0.18\%$
from the result predicted by relation (\ref{advance1}), which is below the
actual error bar.\newline
\indent
This is why we think it is useful to present an alternative way of treating
this problem, which is based on the use of geodesic deviation equations of
first and higher orders. Instead of developing the exact formulae of motion
in terms of powers of the parameter $\frac{GM}{r}$, we propose to start with
an exact solution of a particularly simple form (i.e. a circular orbit with
uniform angular velocity), and then generate the approximate solutions as
geodesics being close to this orbit. \newline
\indent
One of the advantages of this method is the fact that it amounts to treating
consecutively systems of linear equations with constant coefficients, all of
them being of harmonic oscillator type, eventually with an extra right-hand
side being a known periodic function of the proper time. The approximate
solution obtained in this manner has the form of a Fourier series and
represents the closed orbit as a superposition of epicycles with diminishing
amplitude as their circular frequencies grow as multiples of the basic one.
This approach is particularly well-suited for using numerical computations.
An example is provided by the computation of gravitational radiation from
non-circular orbits, for which we use the well-known quadrupole formula.

\section{Geodesic deviations of first and higher orders}

Of many equivalent derivations of geodesic deviation equation we present the
one which most directly leads to the results used in subsequent
applications. Given a (pseudo)-Riemannian manifold $V_{4}$ with the line
element defined by metric tensor $g_{\mu \nu }\,(x^{\lambda })$, 
\begin{equation}
ds^{2}=g_{\mu \nu }\,(x^{\lambda })\,dx^{\mu }dx^{\nu },  \label{line}
\end{equation}
a smooth curve $x^{\lambda }\,(s)$ parametrized with its own length
parameter (or proper time) $s$ is a {\it geodesic} if its tangent vector $%
u^{\mu }=(d\,x^{\mu }/d\,s)$ satisfies the equation: 
\begin{equation}
u^{\lambda }\nabla _{\lambda }u^{\mu }=0\hspace{1em}\Leftrightarrow \hspace{%
1em}\frac{Du^{\mu }}{Ds}=\frac{du^{\mu }}{ds}+\Gamma _{\lambda \rho }^{\mu
}\,u^{\lambda }\,u^{\rho }=0.  \label{geodesic1}
\end{equation}
where $\Gamma _{\rho \lambda }^{\mu }$ denote the Christoffel connection
coefficients of the metric $g_{\mu \nu }$.\newline
\indent
Suppose that a smooth congruence of geodesics is given, of which the
geodesics are labeled by a continuous parameter $p$: $x^{\mu }=x^{\mu
}\,(s,p),$ such that the two independent tangent vector fields are defined
by: 
\begin{equation}
u^{\mu }\,(s,p)=\frac{\partial x^{\mu }}{\partial s}\hspace{1em}{\rm and}%
\hspace{1em}n^{\mu }(s,p)=\frac{\partial x^{\mu }}{\partial p}.
\label{normal}
\end{equation}
It is easily established that the rates of change of the tangent vectors in
the mutually defined directions are equal: 
\begin{equation}
n^{\lambda }\nabla _{\lambda }u^{\mu }=u^{\lambda }\nabla _{\lambda }n^{\mu }%
\hspace{1em}\Leftrightarrow \hspace{1em}\frac{Du^{\mu }}{Dp}=\frac{Dn^{\mu }%
}{Ds}=\frac{\partial ^{2}x^{\mu }}{\partial p\partial s}+\Gamma _{\lambda
\rho }^{\mu }\,\frac{\partial x^{\lambda }}{\partial p}\,\frac{\partial
x^{\rho }}{\partial s},  \label{mixed}
\end{equation}
by virtue of the symmetry of Christoffel symbols in their lower indices. 
\newline
\indent
The Riemann tensor can be defined using covariant derivations along the two
independent directions of the congruence: 
\begin{equation}
\left[ u^{\lambda }\nabla _{\lambda },n^{\rho }\nabla _{\rho }\right] Y^{\mu
}=\Biggl[\frac{D\,}{Ds}\,\frac{D\,}{Dp}-\frac{D\,}{Dp}\,\frac{D\,}{Ds}\Biggr]%
\,Y^{\mu }=R_{\lambda \rho \sigma }^{\;\;\;\;\;\mu }\,\frac{\partial
x^{\lambda }}{\partial s}\,\frac{\partial x^{\rho }}{\partial p}\,Y^{\sigma
}.  \label{Riemann}
\end{equation}
Replacing $Y^{\mu }$ by $u^{\mu }$ in the above formula, we get 
\begin{equation}
\left[ u^{\lambda }\nabla _{\lambda },n^{\rho }\nabla _{\rho }\right] u^{\mu
}=R_{\lambda \rho \sigma }^{\;\;\;\;\;\mu }\,\frac{\partial x^{\lambda }}{%
\partial s}\,\frac{\partial x^{\rho }}{\partial p}\,u^{\sigma }=R_{\lambda
\rho \sigma }^{\;\;\;\;\;\mu }\,u^{\lambda }\,u^{\sigma }\,n^{\rho }.
\label{Riemann2}
\end{equation}
By virtue of the geodesic equation (\ref{geodesic1}) and Eq. (\ref{mixed}),
this can be written as 
\begin{equation}
u^{\lambda }\nabla _{\lambda }\left( n^{\rho }\nabla _{\rho }u^{\mu }\right)
=\frac{D\,}{Ds}\,\frac{Du^{\mu }}{Dp}\,=\frac{D^{2}n^{\mu }}{Ds^{2}}%
=R_{\lambda \rho \sigma }^{\;\;\;\;\;\mu }\,u^{\lambda }\,u^{\sigma
}\,n^{\rho }.  \label{geodev1a}
\end{equation}
This first-order geodesic deviation equation is often called {\it the Jacobi
equation}, and is manifestly covariant.\newline
\indent
In certain applications, Eq.\ (\ref{geodev1a}) can be replaced by its more
explicit, although non-manifestly covariant version: 
\begin{equation}
\frac{d^{2}n^{\mu }}{ds^{2}}+2\,\Gamma _{\lambda \rho }^{\mu }\,u^{\lambda
}\,\frac{dn^{\rho }}{ds}+\partial _{\sigma }\Gamma _{\lambda \rho }^{\mu
}\,u^{\lambda }u^{\rho }n^{\sigma }=0.  \label{geodev1b}
\end{equation}
In this form of the geodesic deviation equation one easily identifies the
relativistic generalizations of the Coriolis-type and centrifugal-type
inertial forces, represented respectively by the second and third terms of
Eq.\ (\ref{geodev1b}).\newline
\indent
The geodesic deviation can be used to construct geodesics $x^{\mu }(s)$
close to a given reference geodesic $x_{0}^{\mu }(s)$, by an iterative
procedure as follows. Let the two geodesics be members of a congruence as
above, with 
\begin{equation}
x^{\mu }(s)=x^{\mu }(s,p),\hspace{2em}x_{0}^{\mu }(s)=x^{\mu }(s,p_{0}).
\label{pert1}
\end{equation}
It follows by direct Taylor expansion, that 
\begin{eqnarray}
x^{\mu }(s,p) &=&{x^{\mu }(s,p_{0})+(p-p_{0})\left. \frac{\partial x^{\mu }}{%
\partial p}\right| _{(s,p_{0})}\,+\frac{1}{2}\,(p-p_{0})^{2}\,\left. \frac{%
\partial ^{2}x^{\mu }}{\partial p^{2}}\right| _{(s,p_{0})}\,+\,...} 
\nonumber \\
&=&{x_{0}^{\mu }(s)+\delta x^{\mu }(s)+\frac{1}{2!}\,\delta ^{2}x^{\mu }(s)+%
\frac{1}{3!}\,\delta ^{3}x^{\mu }(s)+\,...,}  \label{pert2}
\end{eqnarray}
where the more compact notation ${\delta }^{n}{x^{\mu }(s)}$ describes the $%
n^{th}$-order geodesic deviation. Because ${(p-p_{0})}$ is supposed to be a
small quantity, for convenience we may denote it $\epsilon $.\ The
first-order deviation is a vector, ${\delta x^{\mu
}(s)=(p-p_{0})\,n_{0}^{\mu }(s)=\epsilon \,n_{0}^{\mu }(s)}$. But the
second-order deviation is not a vector, and is given by 
\begin{equation}
{\delta }^{2}{x^{\mu }(s)=(p-p_{0})^{2}\,\left( b^{\mu }-\Gamma _{\lambda
\nu }^{\mu }n^{\lambda }n^{\nu }\right) _{0}=\epsilon ^{2}\,\left( b^{\mu
}-\Gamma _{\lambda \nu }^{\mu }n^{\lambda }n^{\nu }\right) _{0}}
\end{equation}
where the covariant second-order deviation vector $b^{\mu }$ is defined by 
\begin{equation}
b^{\mu }=\frac{Dn^{\mu }}{Dp}=\frac{\partial n^{\mu }}{\partial p}%
\,+\,\Gamma _{\lambda \nu }^{\mu }n^{\lambda }n^{\nu }.  \label{pert3}
\end{equation}
Straightforward covariant differentiation of Eq.\ (\ref{geodev1a}), plus use
of the Bianchi and Ricci identities for the Riemann tensor, implies that
this second-order deviation vector $b^{\mu }(s)$ satisfies an inhomogeneous
extension of the first-order geodesic deviation equation: 
\begin{equation}
{\frac{D^{2}b^{\mu }}{Ds^{2}}+R_{\rho \lambda \sigma }^{\;\;\;\;\;\mu
}u^{\lambda }u^{\sigma }b^{\rho }}=[\nabla _{\nu }R_{\lambda \rho \sigma
}^{\;\;\;\;\;\mu }-\nabla _{\lambda }R_{\nu \sigma \rho }^{\;\;\;\;\;\mu
}]u^{\lambda }u^{\sigma }n^{\rho }n^{\nu }+4R_{\lambda \rho \sigma
}^{\;\;\;\;\;\mu }u^{\lambda }n^{\rho }\left( \frac{Dn^{\sigma }}{Ds}\right)
.  \label{geodev2b}
\end{equation}
A more detailed formal derivation of this equation is given in appendix 2.

A rigorous mathematical study of geodesic deviations up to the
second-order, as well as geometric interpretation, but using a different
derivation, was presented in Ref. \cite{Bazanski1}. Also, a
Hamilton--Jacobi formalism has been derived in Refs. \cite{Bazanski2},
which was applied to the problem of free falling particles in the
Schwarzschild space-time \cite{BazansJara}. Fine effects resulting from
the analysis of geodesic deviations of test particles suspended in
hollow spherical satellites have been discussed in Ref. \cite{Shirokov}.

Obviously the procedure can be extended to arbitrarily high order geodesic
deviations ${\delta }^{n}{x^{\mu }(s)}$. This is of considerable practical
importance, as it allows to {\em construct} a desired set of geodesics in
the neighborhood of the reference $x_{0}^{\mu }(s)$, when the congruence of
geodesics is not given a priori in closed form. Indeed, all that is needed
is the set of deviation vectors $(n_{0}^{\mu }(s),b_{0}^{\mu }(s),...)$ {\em %
on the reference geodesic}; obviously these vectors are completely specified
as functions of $s$ by solving the geodesic deviation equations (\ref
{geodev1a}), (\ref{geodev2b}) and their extensions to higher order, for
given $x_{0}^{\mu }(s)$. \newline
\indent
As in the case of the first-order deviation, it is sometimes convenient to
write equation (\ref{geodev2b}) in the equivalent but non-manifest covariant
form 
\begin{equation}
\begin{array}{l}
\displaystyle{\ {\frac{d^{2}b^{\mu }}{ds^{2}}+\partial _{\rho }\,\Gamma
_{\lambda \sigma }^{\mu }\,u^{\lambda }u^{\sigma }b^{\rho }+2\,\Gamma
_{\lambda \sigma }^{\mu }\,u^{\lambda }\,\frac{db^{\sigma }}{ds}=4\left(
\partial _{\lambda }\,\Gamma _{\sigma \rho }^{\mu }+\Gamma _{\sigma \rho
}^{\nu }\,\Gamma _{\lambda \nu }^{\mu }\right) \frac{dn^{\sigma }}{ds}%
\,(u^{\lambda }n^{\rho }-u^{\rho }n^{\lambda })}} \\ 
\\ 
\displaystyle{\ +\left( \Gamma _{\sigma \nu }^{\tau }\,\partial _{\tau
}\,\Gamma _{\lambda \rho }^{\mu }+2\Gamma _{\lambda \tau }^{\mu }\,\partial
_{\rho }\,\Gamma _{\sigma \nu }^{\tau }-\partial _{\nu }\,\partial _{\sigma
}\,\Gamma _{\lambda \rho }^{\mu }\right) (u^{\lambda }u^{\rho }n^{\sigma
}n^{\nu }-u^{\sigma }u^{\nu }n^{\lambda }n^{\rho }).}
\end{array}
\label{geodev2d}
\end{equation}
An equation for the 3rd-order deviation is presented in Appendix 1. \newline
\indent
The non-manifestly covariant geodesic deviation\ equations are often better
adapted to deriving successive approximations for geodesics close to the
initial one. Starting from a given geodesic $x^{\mu }\,(s)$ we can solve
Eq.\ (\ref{geodev1b}) and find the first-order deviation vector $n^{\mu
}\,(s)$. Then, inserting $u^{\mu }\,(s)$ and $n^{\mu }\,(s)$, by now
completely determined, into the system (\ref{geodev2d}), we can solve and
find the second-order deviation vector $b^{\mu }(s)$, and subsequently for
the true second-order coordinate deviation $\delta ^{2}x^{\mu }$, and so
forth. As an example, below we describe non-circular motion, along with
Kepler's laws (in an approximate version), together with the relativistic
perihelion advance, starting from a circular orbit in Schwarzschild metric.%
\newline
\indent 
Although for orbital motion in a Schwarzschild background we have at our
disposal the exact solutions in terms of quadratures (with integrals of
elliptic or Jacobi type), our approach is particularly well-suited for
numerical computations, because in appropriate (Gaussian) coordinates the
geodesic curves can display a very simple parametric form, and all the
components of the $4$-velocity and other quantities reduce to constants when
restricted to that geodesic.\newline
\indent
In this case equation (\ref{geodev1b}) reduces to a linear system with
constant coefficients, which after diagonalization becomes a collection of
harmonic oscillators, and all that remains is to find the characteristic
frequencies. In the next step, we get a collection of harmonic oscillators
excited by external periodic forces represented by the right-hand side of (%
\ref{geodev2d}), which can also be solved very easily, and so forth. \newline
\indent
In the third order, the presence of resonances giving rise to secular terms
could in principle lead to instability of the orbit we started with; but
this phenomenon can be dealt with by Poincar\'{e}'s method \cite{Poincare},
according to which such terms can be eliminated if we admit that the
frequency of the resulting solution is also slightly modified by the
exterior perturbation, and can be expanded in a formal series in successive
powers of the initial (small) deformation parameter.\newline
\indent
At the end, the deviation becomes a series of powers of a small parameter
containing linear combination of characteristic frequencies appearing on the
right-hand side, which are entire multiples of the basic frequency, also
slightly deformed. This description of planetary motion as a superposition
of different harmonic motions has been first introduced by Ptolemaeus in the
II century \cite{Ptolemaios}. We shall now analyse the simplest case of
circular orbits in Schwarzschild geometry.

\section{Circular orbits in Schwarzschild metric}

Let us consider the geodesic deviation equation starting with a circular
orbit in the field of a spherically-symmetric massive body, i.e. in the
Schwarzschild metric. The circular orbits and their stability have been
analyzed and studied in several papers \cite{Droste,Darwin,Sharp} and books,
e.g.\ the well-known monograph by Chandrasekhar \cite{Chandra}.\newline
\indent
The gravitational field is described by the line-element (in natural
coordinates with $c=1$ and $G=1$) 
\begin{equation}
g_{\mu \nu }dx^{\mu }dx^{\nu }=-ds^{2}=-B(r)dt^{2}+\frac{1}{B(r)}
\,dr^{2}+r^{2}\left( d\theta ^{2}+\sin ^{2}\theta \,d\phi ^{2}\right) ,
\label{Schmetric}
\end{equation}
with 
\begin{equation}
B(r)=1-\frac{2M}{r}.  \label{B(r)}
\end{equation}
We recall the essential features of the solution of the geodesic equations
for a test particle of mass $m<<M$. As the spherical symmetry guarantees
conservation of angular momentum, the particle orbits are always confined to
an equatorial plane, which we choose to be the plane $\theta =\pi /2$. The
angular momentum $J$ is then directed along the $z$-axis. Denoting its
magnitude per unit of mass by $\ell =J/m$, we have 
\begin{equation}
\frac{d\phi }{ds}\,=\,\frac{\ell }{r^{2}}\,.  \label{dotphi}
\end{equation}
In addition, as the metric is static outside the horizon $r_{+}=2M$, it
allows a time-like Killing vector which guarantees the existence of a
conserved world-line energy (per unit of mass $m$) $\varepsilon $, such that 
\begin{equation}
\frac{dt}{ds}\,=\,\frac{\varepsilon }{1-\frac{2M}{r}}\,.  \label{energy}
\end{equation}
Finally, the equation for the radial coordinate $r$ can be integrated owing
to the conservation of the world-line Hamiltonian, i.e.\ the conservation of
the absolute four-velocity: 
\begin{equation}
\left( \frac{dr}{ds}\right) ^{2}\,=\,\varepsilon ^{2}\,-\,\left( 1-\frac{2M}{%
r} \right) \left( 1+\frac{\ell ^{2}}{r^{2}}\right) .  \label{conservation}
\end{equation}
From this we derive a simplified expression for the radial acceleration: 
\begin{equation}
\frac{d^{2}r}{ds^{2}}\,=-\,\frac{M}{r^{2}}\,+\left( \frac{\ell ^{2}}{r^{3}}%
\right) \,\left( 1-\frac{3M}{r}\right) .  \label{aradial}
\end{equation}
The equation (\ref{conservation}) can in principle be integrated directly;
indeed, the orbital function $r(\phi )$ is given by an elliptic integral 
\cite{Synge,Weinberg}. However, to get directly an approximate parametric
solution to the equations of motion one can also study {\it perturbations}
of special simple orbits. In the following we study the problem for bound
orbits by considering the first and second-order geodesic deviation
equations for the special case of world lines close to circular orbits.%
\newline
\indent
Observe that for circular orbits $r=R=$ constant, the expressions for $dr/ds$%
, Eq.\ (\ref{conservation}), and $d^{2}r/ds^{2}$, Eq.\ (\ref{aradial}), must
both vanish at all times. This produces two relations between the three
dynamical quantities $(R,\varepsilon ,\ell )$, showing that the circular
orbits are characterized completely by specifying either the radial
coordinate, or the energy, or the angular momentum of the planet. In
particular, the equation for null radial velocity gives 
\begin{equation}
\varepsilon^{2}=\,\left( 1-\frac{2M}{R}\right) \left( 1+\frac{\ell^{2}}{R^{2}%
} \right).  \label{varepsilon}
\end{equation}
Then the null radial acceleration condition (\ref{aradial}) gives the
well-known result 
\begin{equation}
MR^{2}-\ell ^{2}(R-3M)=0\hspace{1em}\Rightarrow \hspace{1em}R\,=\,\frac{\ell
^{2}}{2M}\,\left( 1+\sqrt{\displaystyle{1-\frac{12M^{2}}{\ell ^{2}}}}\right),
\label{wellknown}
\end{equation}
leading to the requirement $R\geq 6M$ for stable circular orbits to exist. 
\newline
\indent
With this in mind, and the explicit formulae for the Christoffel
coefficients of Schwarzschild metric (given in the Appendix 3), we can
establish now the four differential equations that must be satisfied by the
geodesic deviation 4-vector $n^{\mu }\,(s)$ close to a circular orbit. We
recall that on the circular orbit of radius $R$ (which is a geodesic in the
background Schwarzschild metric) we have: 
\begin{equation}
u^{t}=\frac{dt}{ds}=\frac{\varepsilon }{(1-\frac{2M}{R})},\,\,\ \ u^{r}=%
\frac{dr}{ds}=0,\,\ \ u^{\phi }=\frac{d\varphi }{ds}=\omega _{0}=\frac{\ell 
}{R^{2}},\,\ \ u^{\theta }=\frac{d\theta }{ds}=0,  \label{4velocity}
\end{equation}
because $r=R=\mbox{const.}$, $\ \ \theta =\pi /2=\mbox{const.}$, so that $%
\sin \,\theta =1$ \ \ and $\ \ \cos \,\theta =0$.

\section{Geodesic deviation around circular orbit}

It turns out that the four equations are much easier to arrive at if we use
the explicit form of the first-order deviation equation (\ref{geodev1b}). We
get without effort the first three equations, for the components $n^{\theta}$%
, $n^{\phi }$ and $n^{t}$: 
\begin{equation}
\frac{d^{2}n^{\theta }}{ds^{2}}=-\,(u^{\phi })^{2}\,n^{\theta }=-\frac{\ell
^{2}}{R^{4}}\,n^{\theta }\,,  \label{ntheta}
\end{equation}
\begin{equation}
\frac{d^{2}n^{\phi }}{ds^{2}}=-\frac{2\ell }{R^{3}}\,\frac{dn^{r}}{ds}%
,\,\,\, \frac{d^{2}n^{t}}{ds^{2}}=-\frac{2M\varepsilon }{R^{2}(1-\frac{2M}{R}%
)^{2}}\, \frac{dn^{r}}{ds}.  \label{nphi+nt1}
\end{equation}
\indent
The deviation $n^{\theta }$ is independent of the remaining three variables $%
n^{t}$, $n^{r}$ and $n^{\varphi }$. The harmonic oscillator equation (\ref
{ntheta}) for $n^{\theta }$ displays the frequency which is equal to the
frequency of the circular motion of the planet itself: 
\begin{equation}
n^{\theta }\,(s)=n_{0}^{\theta }\,\cos (\omega
_{0}\,s+\gamma)=n_{0}^{\theta}\, \cos \left( \frac{\ell }{R^{2}}s+\gamma
\right).
\end{equation}
This can be interpreted as the result of a change of the coordinate system,
with a new $z$-axis slightly inclined with respect to the original one, so
that the plane of the orbit does not coincide with the plane $z=0$. In this
case the deviation from the plane will be described by the above solution,
i.e.\ a trigonometric function with the period equal to the period of the
planetary motion. Being a pure coordinate effect, it allows us to eliminate
the variable $n^{\theta }$ by choosing $n^{\theta }=0$. \newline
\indent 
It takes a little more time to establish the equation for $n^{r}$, using
Eq.\ (\ref{geodev1b}): 
\begin{equation}
\frac{d^{2}n^{r}}{ds^{2}}+2\,\Gamma _{\lambda \rho }^{r}\,u^{\lambda }\, 
\frac{dn^{\rho }}{ds}+\partial _{\sigma }\Gamma _{\lambda \rho}^{r}\,
u^{\lambda }u^{\rho }n^{\sigma }=0.  \label{nra}
\end{equation}
Taking into account that only the components $u^{t}$ and $u^{\phi }$ of the
four-velocity on the circular orbit are different from zero, and recalling
that we have chosen to set $n^{\theta }=0$, too, the only non-vanishing
terms in the above equation are: 
\begin{equation}
\frac{d^{2}n^{r}}{ds^{2}}+2\,\Gamma _{tt}^{r}\,u^{t}\,\frac{dn^{t}}{ds}
+2\,\Gamma _{\phi \phi }^{r}\,u^{\phi }\,\frac{dn^{\phi }}{ds} +
\partial_{r} \Gamma_{tt}^{r}\,u^{t}u^{t}n^{r}+\partial _{r}\Gamma_{\phi
\phi}^{r}\, u^{\phi }u^{\phi }n^{r}=0.  \label{nrb}
\end{equation}
Using the identities (\ref{wellknown}) and the definitions (\ref{4velocity}%
), we get 
\begin{equation}
\frac{d^{2}n^{r}}{ds^{2}}-\frac{3\ell ^{2}}{R^{4}}\,\biggl(1-\frac{2M}{R} %
\biggr)\,n^{r}+\frac{2M\varepsilon }{R^{2}}\,\frac{dn^{t}}{ds}-\frac{2\ell}{
R}\,\biggl(1-\frac{2M}{R}\biggr)\,\frac{dn^{\varphi }}{ds}=0.  \label{nr1}
\end{equation}
The system of three remaining equations can be expressed in a matrix form: 
\begin{equation}
\pmatrix{\frac{d^2}{ds^2} & \frac{2M \varepsilon}{R^2 (1 -
\frac{2M}{R})^2}\, \frac{d \,}{ds} & 0 \cr \frac{2M \varepsilon}{R^2}
\frac{d \,}{ds} & \frac{d^2}{ds^2} - \frac{3 \ell^2}{R^4} \, ( 1 -
\frac{2M}{R} ) & - \frac{2 \ell}{R} \, ( 1 - \frac{2M}{R} ) \,
\frac{d\,}{ds} \cr 0 & \frac{2 \ell}{R^3} \, \frac{d \,}{ds} &
\frac{d^2}{ds^2} } \,\,\pmatrix{n^t \cr n^r \cr n^{\varphi} }\,=\,\pmatrix{0
\cr 0 \cr 0}.  \label{matrixn}
\end{equation}
The characteristic equation of the above matrix is 
\begin{equation}
\lambda ^{4}\,\biggl[\lambda ^{2}+\frac{\ell ^{2}}{R^{4}}\,\biggl(1-\frac{2M 
}{R}\biggr)-\frac{4M\varepsilon ^{2}}{R^{4}(1-\frac{2M}{R})^{2}}\biggr]=0,
\label{omega1}
\end{equation}
which after using the identities (\ref{varepsilon}) and (\ref{wellknown})
reduces to 
\begin{equation}
\lambda^{4}\,\biggl[\lambda ^{2}+\frac{\ell ^{2}}{R^{4}}\,\biggl(1-\frac{6M}{%
R} \biggr)\biggr]=0,  \label{omega2}
\end{equation}
so that the characteristic circular frequency is 
\begin{equation}
\omega =\frac{\ell }{R^{2}}\,\sqrt{1-\frac{6M}{R}}=\omega _{0}\,\sqrt{1- 
\frac{6M}{R}}.  \label{omega3}
\end{equation}
It is obvious that the general solution contains oscillating terms $\cos
(\omega s)$ ; however, before we analyse in detail this part of solution,
let us consider the terms linear in the variable $s$ or constants: as a
matter of fact, because of the presence of first and second-order
derivatives with respect to $s$ in the matrix operator (\ref{matrixn}), the
general solution may also contain the following vector: 
\begin{equation}
\pmatrix{ (\Delta \, u^t ) \, s + \Delta \, t \cr ( \Delta \, u^r) \, s +
\Delta \, r \cr ( \Delta \, u^{\varphi} ) \, s + \Delta \, \varphi }.
\label{linear}
\end{equation}
When inserted into the system (\ref{matrixn}), the solution is the
following: \vskip0.2cm \indent
\hskip0.5cm $\Delta \,t$ \thinspace\ and \thinspace\ $\Delta \,\varphi$
\thinspace\ are arbitrary; \vskip0.1cm \indent
\hskip0.5cm $\Delta \,u^{r}=0$, which means that the radial velocity remains
null; and 
\begin{equation}
\frac{3\ell^{2}}{R^{4}}\,\biggl(1-\frac{2M}{R}\biggr)\, \Delta\, r= \frac{%
2M\varepsilon}{R^{2}}\,\Delta \,u^{t}-\frac{2\ell }{R}\, \biggl(1-\frac{2M}{R%
}\biggr)\,\Delta \,u^{\varphi }=0.  \label{gauge}
\end{equation}
This condition coincides with the transformation of the initial circular
geodesic of radius $R$ to a neighboring one, with radius $R+\Delta r$, with
the subsequent variations $\Delta \,u^{t}$ and $\Delta u^{\varphi }$ added
to the corresponding components of the 4-velocity in order to satisfy the
condition $g_{\mu \nu }\,u^{\mu }u^{\nu }=1$ in the linear approximation.
After choosing an optimal value for $r$, we can forget about this particular
solution, as well as about the arbitrary shift in the variables $t$ and $%
\phi $, and investigate the oscillating part of the solution.\newline
\indent
We shall choose the initial phase to have (with $n_{0}^{r}>0)$: 
\begin{equation}
n^{r}\,(s)=-n_{0}^{r}\,\cos \,(\omega s).
\end{equation}
What remains to be done is to compare this frequency with the fundamental
circular frequency $\omega_{0}=\ell /R^{2}$ of the unperturbed circular
orbital motion.\newline
\indent
But this discrepancy between the two circular frequencies $\omega$ and $%
\omega _{0}$ is exactly what produces the perihelion advance, and its value
coincides with the value obtained in the usual way (\ref{advance1}) in the
limit of quasi-circular orbits, i.e.\ when $e^{2}\rightarrow 0$: we get both
the correct value and the correct sign.\newline
\indent
Let us display the complete solution for the first-order deviation vector $%
n^{\mu}\,(s)$ which takes into account only the non-trivial degrees of
freedom: 
\begin{equation}
n^{\theta }=0,\,n^{r}(s)=-n_{0}^{r}\,\cos (\omega\,s),\,\, n^{\varphi
}=n_{0}^{\varphi }\,\sin (\omega \,s),\,\, n^{t}=n_{0}^{t}\,\sin (\omega
\,s).  \label{n}
\end{equation}
The only independent amplitude is given by $n_{0}^{r}$, because we have 
\begin{eqnarray}
n_{0}^{t} &=&\frac{2M\varepsilon }{R^{2}(1-\frac{2M}{R})^{2}\omega }
n_{0}^{r}=\frac{2\sqrt{M}}{\sqrt{R}\left( 1-\frac{2M}{R}\right) \sqrt{1- 
\frac{6M}{R}}}n_{0}^{r}\,,  \label{n0t} \\
n_{0}^{\varphi } &=&\frac{2\ell }{R^{3}\omega }\,n_{0}^{r}=\frac{2\omega
_{0} }{R\omega }\,n_{0}^{r}=\frac{2}{R\sqrt{1-\frac{6M}{R}}}\,n_{0}^{r}\,.
\label{n0varphi}
\end{eqnarray}
So the trajectory and the law of motion are given by 
\begin{eqnarray}
r &=&R-\,n_{0}^{r}\,\cos (\omega s),  \label{r1order} \\
\varphi &=&\omega _{0}\,s+\,n_{0}^{\varphi }\,\sin (\omega \,s)=\frac{\sqrt{%
M }}{R^{3/2}\sqrt{1-\frac{3M}{R}}}s+\,n_{0}^{\varphi }\,\sin (\omega \,s),
\label{varphi1order} \\
t &=&\frac{\varepsilon }{(1-\frac{2M}{R})}s+\,n_{0}^{t}\,\sin (\omega \,s)= 
\frac{1}{\sqrt{1-\frac{3M}{R}}}s+\,n_{0}^{t}\,\sin (\omega \,s),
\label{t1order}
\end{eqnarray}
where the phase in the argument of the $cosine$ function was chosen so that $%
s=0$ corresponds to the perihelion, and $s=\frac{\pi }{\omega }$ to the
aphelion. It is important to note once again that the coefficient $n_{0}^{r}$%
, which also fixes the values of the two remaining amplitudes, $n_{0}^{t}$
and $n_{0}^{\varphi }$, defines the size of the actual deviation, so that
the ratio $\frac{n_{0}^{r}}{R}$ becomes the dimensionless infinitesimal
parameter controlling the approximation series with consecutive terms
proportional to the consecutives powers of $\frac{n_{0}^{r}}{R}$. \newline
\indent
What we see here is the approximation to an elliptic orbital movement as
described by the presence of an {\it epicycle} (exactly like in the
Ptolemean system \cite{Ptolemaios}, except for the fact that the Sun is
placed in the center instead of the Earth). As a matter of fact, the
development into power series with respect to the eccentricity $e$
considered as a small parameter, and truncating all the terms except the
linear one, leads to the Kepler result \cite{Kepler}, 
\begin{equation}
r(t)=\frac{a(1-e^{2})}{1+e\,\cos (\omega _{0}\,t)}\simeq a\,\left[ 1-e\,\cos
(\omega _{0}\,t)\right] ,  \label{ellipse}
\end{equation}
which looks almost as our formula (\ref{r1order}) if we identify the
eccentricity $e$ with $\frac{n_{0}^{r}}{R}$ and the greater half-axis $a$
with $R$; but there is also the additional difference, that the circular
frequency of the epicycle is now slightly lower than the circular frequency
of the unperturbed circular motion.\newline
\indent
But if the circular frequency is lower, the period is slightly longer: in a
linear approximation, we have 
\begin{equation}
\omega =\sqrt{\frac{\ell ^{2}}{R^{4}}\,\left( 1-\frac{6M}{R}\right) }\,,
\label{omegalin}
\end{equation}
hence keeping the terms up to the third order in $\frac{M}{R}$, 
\begin{equation}
T\simeq T_{0}\,\left( 1+\frac{3M}{R}+\frac{27}{2}\,\frac{M^{2}}{R^{2}}+\frac{
135}{2}\,\frac{M^{3}}{R^{3}}+...\right) .
\end{equation}
Then obviously one must have $\frac{\Delta \varphi }{2\pi }=\frac{\Delta T}{
T_{0}}$ from which we obtain the perihelion advance after one revolution 
\begin{equation}
\Delta \varphi =\frac{6\,\pi M}{R}+\frac{27\,\pi M^{2}}{R^{2}}+\frac{
135\,\pi M^{3}}{R^{3}}+...  \label{advance4}
\end{equation}
\indent
It is obvious that at this order of approximation we could not keep track of
the factor $(1-e^{2})^{-1}$, containing the eccentricity (here replaced by
the ratio $\frac{n_{0}^{r}}{R}$) only through its square. In contrast, we
obtain without effort the coefficients in front of terms quadratic or cubic
in $\frac{M}{R}$. This shows that our method can be of interest when one has
to consider the low-eccentricity orbits in the vicinity of very massive and
compact bodies, having a non-negligible ratio $\frac{M}{R}$.\newline
\indent
In order to include this effect, at least in its approximate form as the
factor $(1+e^{2})$, we must go beyond the first-order deviation equations
and investigate the solutions of the equations describing the quadratic
effects (\ref{geodev2d}).

\section{The second-order geodesic deviation}

After inserting the complete solution for the first-order deviation vector (%
\ref{n})--(\ref{n0varphi}) into the system (\ref{geodev2b}) and a tedious
calculation, we find the following set of linear equations satisfied by the
second-order deviation vector $b^{\mu }(s)$: 
\begin{equation}
\pmatrix{\frac{d^2}{ds^2} & \frac{2M \varepsilon}{R^2 (1 -
\frac{2M}{R})^2}\, \frac{d \,}{ds} & 0 \cr \frac{2M \varepsilon}{R^2}
\frac{d \,}{ds} & \frac{d^2}{ds^2} - \frac{3 \ell^2}{R^4} \, ( 1 -
\frac{2M}{R} ) & - \frac{2 \ell}{R} \, ( 1 - \frac{2M}{R} ) \, \frac{d
\,}{ds} \cr 0 & \frac{2 \ell}{R^3} \, \frac{d \,}{ds} & \frac{d^2}{ds^2}}\,\,%
\pmatrix{b^t \cr b^r \cr b^{\varphi} }=(n_{0}^{r})^{2}\,\pmatrix{C^t \cr C^r
\cr C^\varphi}\,,  \label{matrixb}
\end{equation}
where we have put into evidence the common factor $(n_{0}^{r})^{2}$, which
shows the explicit quadratic dependence of the second-order deviation vector 
$b^{\mu }$ on the first-order deviation amplitude $n_{0}^{r}$. The constants 
$C^{t},C^{r}$ and $C^{\varphi }$ are expressions depending on $M$, $R$, $%
\omega _{0}$, $\omega $, $\varepsilon $, $\sin (2\omega s)$ and $\cos
(2\omega s)$: 
\begin{equation}
C^{t}=-\frac{6M^{2}(2-\frac{7M}{R})\varepsilon \sin \left( 2\omega s\right) 
}{(1-\frac{3M}{R})(1-\frac{2M}{R})^{2}R^{6}\omega }\,,  \label{Ct}
\end{equation}
\begin{equation}
C^{r}=\frac{3M\left[ (2-\frac{5M}{R}+\frac{18M^{2}}{R^{2}})-(6-\frac{27M}{R}+%
\frac{6M}{R^{2}}^{2})\cos \left( 2\omega s\right) \right] }{2(1-\frac{3M}{R}%
)(1-\frac{6M}{R})R^{4}}\,,  \label{Cr}
\end{equation}
\begin{equation}
C^{\varphi }=-\frac{6M(1-\frac{M}{R})\omega _{0}\sin \left( 2\omega s\right) 
}{(1-\frac{3M}{R})R^{5}\omega }\,.  \label{Cvarphi}
\end{equation}
\indent
The solution of the above matrix for $b^{\mu }(s)$ has the same
characteristic equation of the matrix (\ref{matrixn}) for $n^{\mu }(s)$, and
the general solution containing oscillating terms with\ angular frequency $%
\omega $ is of no interest because it is already accounted for by $n^{\mu
}(s)$. But the particular solution includes the terms linear in the proper
time $s$, constant ones, and the terms oscillating with\ angular frequency $%
2\omega $: 
\begin{equation}
b^{t}=\frac{\left( n_{0}^{r}\right) ^{2}M\varepsilon }{R^{3}(1-\frac{6M}{R}%
)(1-\frac{2M}{R})^{2}}\left[ -\frac{3(2-\frac{5M}{R}+\frac{18M^{2}}{R^{2}})}{%
1-\frac{6M}{R}}s+\frac{2-\frac{13M}{R}}{\omega }\sin (2\omega s)\right] ,
\label{bt}
\end{equation}
\begin{equation}
b^{r}=\frac{\left( n_{0}^{r}\right) ^{2}}{2R(1-\frac{6M}{R})}\left[ \frac{%
3(2-\frac{5M}{R}+\frac{18M^{2}}{R^{2}})}{1-\frac{6M}{R}}+\left( 2+\frac{5M}{R%
}\right) \cos \left( 2\omega s\right) \right] ,  \label{br}
\end{equation}
\begin{equation}
b^{\varphi }=\frac{\left( n_{0}^{r}\right) ^{2}\omega _{0}}{R^{2}(1-\frac{6M%
}{R})}\left[ -\frac{3(2-\frac{5M}{R}+\frac{18M^{2}}{R^{2}})}{1-\frac{6M}{R}}%
s+\frac{1-\frac{8M}{R}}{2\omega }\sin (2\omega s)\right] .  \label{bvarphi}
\end{equation}
\indent
As explained in Section $2$, we need to calculate $\frac{1}{2}\,\delta
^{2}x^{\mu }$ to obtain the geodesic curve $x^{\mu }$ with second-order
geodesic deviation: 
\begin{equation}
\delta ^{2}t=\frac{\left( n_{0}^{r}\right) ^{2}M\varepsilon }{R^{3}}\left[ -%
\frac{3(2-\frac{5M}{R}+\frac{18M}{R^{2}}^{2})}{(1-\frac{2M}{R})^{2}(1-\frac{%
6M}{R})^{2}}s+\frac{2-\frac{15M}{R}+\frac{14M^{2}}{R^{2}}}{(1-\frac{6M}{R}%
)(1-\frac{2M}{R})^{3}\omega }\sin (2\omega s)\right] ,  \label{d2t}
\end{equation}
\begin{equation}
\delta ^{2}r=\frac{\left( n_{0}^{r}\right) ^{2}}{R(1-\frac{6M}{R})}\left[ 
\frac{5-\frac{33M}{R}+\frac{90M^{2}}{R^{2}}-\frac{72M^{3}}{R^{3}}}{(1-\frac{%
2M}{R})(1-\frac{6M}{R})}-\left( 1-\frac{7M}{R}\right) \cos (2\omega s)\right]
,  \label{d2r}
\end{equation}
\begin{equation}
\delta ^{2}\varphi =\frac{\left( n_{0}^{r}\right) ^{2}\omega _{0}}{R^{2}(1-%
\frac{6M}{R})}\left[ -\frac{3(2-\frac{5M}{R}+\frac{18M^{2}}{R^{2}})}{1-\frac{%
6M}{R}}s+\frac{5-\frac{32M}{R}}{2\omega }\sin (2\omega s)\right] .
\label{d2varphi}
\end{equation}
The fact that the second-order deviation vector $b^{\mu }$ turns with
angular frequency $2\omega $ enables us to get a better approximation of the
elliptic shape of the resulting orbit. The trajectory described by $x^{\mu }$
including second-order deviations is not an ellipse, but we can match the
perihelion and aphelion distances to see that $R\neq a$ and $e\neq
n_{0}^{r}/R$ when second-order deviation is used. The perihelion and
aphelion distances of the Keplerian, i.e., elliptical orbit are $a(1-e)$ and 
$a(1+e)$. For $x^{\mu }$, the perihelion is obtained when $\omega s=2k\pi $
and the aphelion when $\omega s=(1+2k)\pi $, where $k\in {\Bbb Z}$. Matching
the radius for perihelion and aphelion, we obtain the semimajor axis $a$ and
the eccentricity $e$ of an ellipse that has the same perihelion and aphelion
distances of the orbit described by $x^{\mu }$: 
\begin{equation}
a=R+\frac{\left( n_{0}^{r}\right) ^{2}}{12R}\left[ -1+\frac{3}{1-\frac{2M}{R}%
}+\frac{7}{1-\frac{6M}{R}}+\frac{15}{(1-\frac{6M}{R})^{2}}\right] 
\label{a2order}
\end{equation}
\begin{equation}
e=\frac{n_{0}^{r}(1-\frac{2M}{R})(1-\frac{6M}{R})^{2}}{R(1-\frac{2M}{R})(1-%
\frac{6M}{R})^{2}+\frac{\left( n_{0}^{r}\right) ^{2}}{R}\left[ 2-\frac{9M}{R}%
+\frac{11M^{2}}{R^{2}}+\frac{6M^{3}}{R^{3}}\right] }=\frac{n_{0}^{r}}{R}+%
{\cal O}\left( \frac{(n_{0}^{r})^{3}}{R^{3}}\right) .  \label{e2order}
\end{equation}
In the limit case of $\frac{M}{R}\rightarrow 0$, there is no perihelion
advance and $a=R\left[ 1+2(\frac{n_{0}^{r}}{R})^{2}\right] $ and $e=\frac{%
n_{0}^{r}}{R}$, so the second-order deviation increases the semimajor axis $a
$ of a matching ellipse compared to the first-order deviation, when $a=R$
and $e=\frac{n_{0}^{r}}{R}$.\newline
\indent
Another comparison with elliptic orbits concerns the shape of the orbit
described by $r(\varphi )$. From $\varphi (s)$ it is possible to write $%
s(\varphi )$ by means of successive approximations, beginning with $\omega
s=\varphi \sqrt{1-\frac{6M}{R}}$. Finally, $s$ can be replaced in $r(s)$ and
we obtain $r(\varphi )$ up to the second order in $\frac{n_{0}^{r}}{R}$: 
\begin{eqnarray}
\frac{r}{R} &=&1-\frac{n_{0}^{r}}{R}\cos \left( \frac{\omega }{\omega _{0}}%
\varphi \right) +\left( \frac{n_{0}^{r}}{R}\right) ^{2}\left[ \frac{3-\frac{%
5M}{R}-\frac{30M^{2}}{R^{2}}+\frac{72M^{3}}{R^{3}}}{2(1-\frac{2M}{R})(1-%
\frac{6M}{R})^{2}}\right.  \\
&&\left. +\frac{(1-\frac{5M}{R})}{2(1-\frac{6M}{R})}\cos \left( \frac{%
2\omega }{\omega _{0}}\varphi \right) \right] +...  \label{r2order}
\end{eqnarray}
In the limit $\frac{M}{R}\rightarrow 0$, the exact equation of a ellipse is
obtained up to the second order in $e$, where $e=n_{0}^{r}/R$ and $%
r_{0}=(1+e^{2})R$: 
\begin{equation}
r=\frac{r_{0}}{1+e\,\cos \varphi }=\frac{(1+e^{2})R}{1+e\,\cos \varphi }=R%
\left[ 1-e\cos \varphi +e^{2}\left( \frac{3}{2}+\frac{1}{2}\cos 2\varphi
\right) +...\right] .  \label{ellipser0}
\end{equation}
Comparing with the ellipse equation (\ref{ellipse}), we have $%
r_{0}=a(1-e^{2})$, so $a=\frac{R(1+e^{2})}{(1-e^{2})}\simeq R(1+2e^{2})$
which agrees with the analysis of Eqs. (\ref{a2order})--(\ref{e2order}).

\section{Third-order terms and Poincar\'{e}'s method}

With the third-order approximation we are facing a new problem, arising from
the presence of {\it resonance terms} on the right-hand side. It is easy to
see that after reducing the expressions on the right-hand side of equation (%
\ref{geodev3b}) in Appendix 1, which contain the terms of the form 
\[
\cos^{3}\,\omega s,\,\ \ \,\ \,\sin \,\omega s\,\cos ^{2}\,\omega s 
\]
and the like, we shall get not only the terms containing 
\[
\sin \,3\,\omega \,s\,,\,\ \ \,{\rm and}\,\ \ \,\cos \,3\,\omega \,s\,\, 
\]
which do not create any particular problem, but also the {\it resonance terms%
} containing the functions $\sin \,\omega \,s$ and $\cos \,\omega \,s$ ,
whose circular frequency is the same as the eigenvalue of the
matrix-operator acting on the left-hand side.

As a matter of fact, the equation for the covariant third-order deviation $%
h^{\mu }$ can be written in matrix form, with principal part linear in the
third-order deviation $h^{\mu }$, represented by exactly the same
differential operator as in the lower-order deviation equations. The
right-hand side is separated into two parts, one oscillating with frequency $%
\omega $, and another with frequency $3\omega $: 
\begin{equation}
\begin{array}{l}
{\displaystyle\pmatrix{\frac{d^2}{ds^2} & \frac{2M \varepsilon}{R^2 (1 -
\frac{2M}{R})^2}\, \frac{d \,}{ds} & 0 \cr \frac{2M \varepsilon}{R^2}
\frac{d \,}{ds} & \frac{d^2}{ds^2} - \frac{3 \ell^2}{R^4} \, ( 1 -
\frac{2M}{R} ) & - \frac{2 \ell}{R} \, ( 1 - \frac{2M}{R} ) \, \frac{d
\,}{ds} \cr 0 & \frac{2 \ell}{R^3} \, \frac{d \,}{ds} & \frac{d^2}{ds^2} }%
\,\,\pmatrix{h^t \cr h^r \cr h^{\varphi} }=} \\ 
\\ 
{\displaystyle\hspace{2em}=(n_{0}^{r})^{3}\,\pmatrix{B^t \, \sin ( \omega s)
+ C^t \, \sin (\, 3 \omega s) + s\, D^t \, \cos ( \omega s)\cr B^r \,
\cos(\omega s) + C^r \, \cos (\, 3 \omega s) + s\,D^r \, \sin( \omega s) \cr
B^{\varphi} \, \sin (\omega s) + C^{\varphi} \sin (\, 3 \omega s)+
s\,D^{\varphi} \, \cos (\omega s) },}
\end{array}
\label{matrix3}
\end{equation}
where the coefficients $B^{k}$, $C^{k}$ and $D^{k}$, $k=t,r,\varphi $ are
complicated functions of $\frac{M}{R}$.

The proper frequency of the matrix operator acting on the left-hand side is
equal to $\omega $; the terms containing the triple frequency $3\,\omega $
will give rise to the unique non-singular solution of the same frequency,
but the resonance terms of the basic frequency on the right-hand side will
give rise to secular terms, proportional to $s$, which is in contradiction
with the bounded character of the deviation we have supposed from the
beginning. The term proportional to $s$ on the right-hand side is eliminated
in the differential equation for $h^{r}$ when $\frac{dh^{\varphi }}{ds}$ and 
$\frac{dh^{t}}{ds}$ are replaced by theirs values.

Poincar\'{e} \cite{Poincare}\ was first to understand that in order to solve
this apparent contradiction, one has to take into account possible
perturbation of the basic frequency itself, which amounts to the replacement
of $\omega $ by an infinite series in powers of the infinitesimal parameter,
which in our case is the eccentricity $e=\frac{{n_{0}^{r}}}{R}$: 
\begin{equation}
\omega \rightarrow \omega +e\,\omega _{1}+e^{2}\,\omega _{2}+e^{3}\,\omega
_{3}+\dots \,,  \label{newomega}
\end{equation}
Then, developing both sides into a series of powers of the parameter $e$, we
can not only recover the former differential equations for the vectors $%
n^{\mu },\,b^{\mu },\,h^{\mu }$, but get also some algebraic relations
defining the corrections $\omega _{1},\,\omega _{2},\,\omega _{3},$ etc. 
\newline
\indent
The equations resulting from the requirement that all resonant terms on the
right-hand side be canceled by similar terms on the left-hand side are
rather complicated. We do not attempt to solve them here. However, one
easily observes that the absence of resonant terms in the second-order
deviation equations forces $\omega _{1}$ to vanish, while the next term $%
\omega _{2}$ is different from $0.$\newline
\indent
Similarly, as there are no resonant terms in the equations determining the
fourth-order deviation, because all four-power combinations of sine and
cosine functions will produce terms oscillating with frequencies $2\omega $
and $4\omega $; as a result, the correction $\omega _{3}$ will be also equal
to $0$. Next secular terms will appear at the fifth-order approximation, as
products of the type $cos^{5}\omega s$, $sin^{3}\omega s\,cos^{2}\omega s$,
etc, produce resonant terms again, which will enable us to find the
correction $\omega _{4}$, and so on, so that the resulting series
representing the frequency $\omega $ contains only even powers of the small
parameter $\frac{n_{0}^{r}}{R}$.

\section{Gravitational radiation}

The decomposition of the elliptic trajectory turning slowly around its focal
point into a series of epicycles around a circular orbit can also serve for
obtaining an approximate spectral decomposition of gravitational waves
emitted by a celestial body moving around a very massive attracting center.%
\newline
\indent
It is well known that gravitational waves are emitted when the quadrupole
moment of a mass distribution is different from zero, and the amplitude of
the wave is proportional to the third derivative of the quadrupole moment
with respect to time (in the reference system in which the center of mass
coincides with the origin of the Cartesian basis in three dimensions, see
Ref. \cite{Landau}).\newline
\indent
Of course, it is only a linear approximation, but it takes the main features
of the gravitational radiation emitted by the system well into account,
provided the velocities and the gravitational fields are not relativistic
and the wavelength of gravitational radiation is large compared to the
dimensions of the source (quadrupole approximation).\newline
\indent
More precisely, let us denote the tensor $Q_{ij}$ of a given mass
distribution $\mu \,(x_{i}),$ where $i,j,=1,2,3$, see Ref. \cite{Peters}: 
\begin{equation}
Q_{ij}=\int \,\mu \,\,x_{i}\,x_{j}dV\,=\sum_{\alpha }m_{\alpha }x_{\alpha
i}\, x_{\alpha j},  \label{Qtensor}
\end{equation}
where $m_{\alpha }$ are point masses.\newline
\indent
Let $\overrightarrow{OP}$ be the vector pointing at the observer (placed at
the point $P$), from the origin of the coordinate system coinciding with the
center of mass of the two orbiting bodies whose motion is approximately
described by our solution in a Fourier series form. It is also supposed that
the length of this vector is much greater than the characteristic dimensions
of the radiating system, i.e. $\mid \overrightarrow{OP}\mid \gg R$.\newline
\indent
Then the total power of gravitational radiation $P$ emitted by the system
over all directions is given by the following expression (see Ref. \cite
{Peters}): 
\begin{equation}
P=\frac{G}{5c^{5}}\,\,\biggl(\frac{d^{3}Q_{ij}}{dt^{3}}\,\frac{d^{3}Q_{ij}}{
dt^{3}}-\frac{1}{3}\,\frac{d^{3}Q_{ii}}{dt^{3}}\,\frac{d^{3}Q_{jj}}{dt^{3}} %
\biggr)\,.  \label{P}
\end{equation}
\newline
\indent
When applied to Keplerian motion of two masses $m_{1}$ and $m_{2}$, with
orbit equation and angular velocity given by 
\begin{equation}
r=\frac{a(1-e^{2})}{1+e\,\cos \varphi }\,,\,\ \ \,\ \frac{d\varphi }{dt}= 
\frac{\sqrt{G(m_{1}+m_{2})a(1-e^{2})}}{r^{2}}\,,  \label{KeplerOrbit}
\end{equation}
the total power $P$ now reads 
\begin{equation}
P=\frac{8}{15}\,\frac{G^{4}}{c^{5}}\,\frac{m_{1}^{2}m_{2}^{2}(m_{1}+m_{2})}{
a^{5}(1-e^{2})^{5}}\,(1+e\,\cos \varphi )^{4}\left[ 12\,(1+e\,\cos \varphi
)^{2}+e^{2}\,\sin ^{2}\varphi \right] .  \label{PKepler}
\end{equation}
\newline
\indent
We shall calculate the $P$ in Eq.\ (\ref{P}) with our solution $x^{\mu }$
using second-order geodesic deviation, to inspect the non-negligible effects
of the ratio $\frac{M}{R}$. We have the explicit solutions $r(s)$, $\varphi
(s)$ and $t(s)$, so to calculate $\frac{dQ_{ij}}{dt}$ we need only the
derivatives with respect to $s$, i.e., $\frac{df}{dt}=\frac{df}{ds}/\frac{dt%
}{ds}$ can be applied successively to obtain $\frac{d^{3}Q_{ij}}{dt^{3}}$.
So we finally get $P$ as function of $s$, which is not shown here because it
is a very large expression that nevertheless can be easily obtained using a
symbolic calculus computer program. \newline
\indent
As we want to compare the two total powers $P$ during one orbital period
(between perihelions), $P$ in the Kepler case is obtained from the numerical
solution for $\varphi (t)$ calculated from Eq. (\ref{KeplerOrbit}), and $P$
of the geodesic deviation case has to use $s(t)$ obtained from $t(s)$ by
means of successive approximations, starting with $s=\frac{t}{\varepsilon } 
\sqrt{1-\frac{2M}{R}}$. \newline
\indent
There are many possible ways to compare a Keplerian orbit with a
relativistic one. Here we assume $m_{1}\gg m_{2}$ and fix the values of $a$, 
$e$, $m_{1}$; the values of $R$ and $n_{0}^{r}$ are calculated to obtain an
exact ellipse (up to the second order in $e$) in the limit $\frac{M}{R}%
\rightarrow 0$, like Eq. (\ref{ellipser0}),\ so $R=\frac{(1-e^{2})}{(1+e^{2})%
}a$ and $n_{0}^{r}=R\ e$. Up to first order in $e$, we have $R=a$. The
choice of $M=m_{1}$ allows the two total powers $P$ to be equal when $e=0$
and $\frac{M}{R}\rightarrow 0$. Figures $1$ and $2$ show this comparison for
small eccentricities and non-negligible $\frac{M}{R}$ ratios.

\begin{figure}[t]
\begin{center}
\includegraphics[scale=0.80]{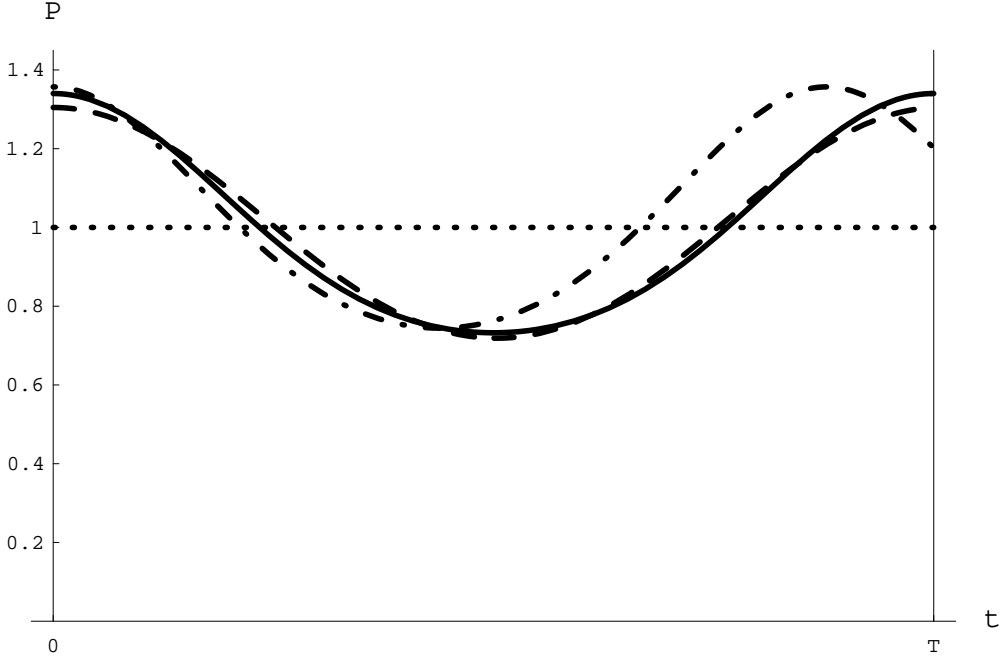}
\end{center}
\caption{{\protect\footnotesize The total power $P$ in four cases as
function of $t$ during one orbital period $T$, with $M=m_{1}$. The dotted
line is the circular orbit, i.e., $e=n_{0}^{r}=0$ case, and its total power $%
P$ is used as a reference to the others. With $e=0.05$ and $\frac{m_{1}}{a}
=0.04$, the total power $P$ for elliptic Keplerian orbit is represented by
the dot-dashed line. The dashed line is $P$ with $x^{\protect\mu }$ with
first-order geodesic deviation, and $\frac{n_{0}^{r}}{R}=0.05$ and $\frac{M}{
R}=0.04$. Finally, the total power $P$ with second-order geodesic deviation
is given by the solid line, where $\frac{n_{0}^{r}}{R}=0.05$ and $\frac{M}{R}%
=0.0402$.}}
\end{figure}

\begin{figure}[t]
\begin{center}
\includegraphics[scale=0.80]{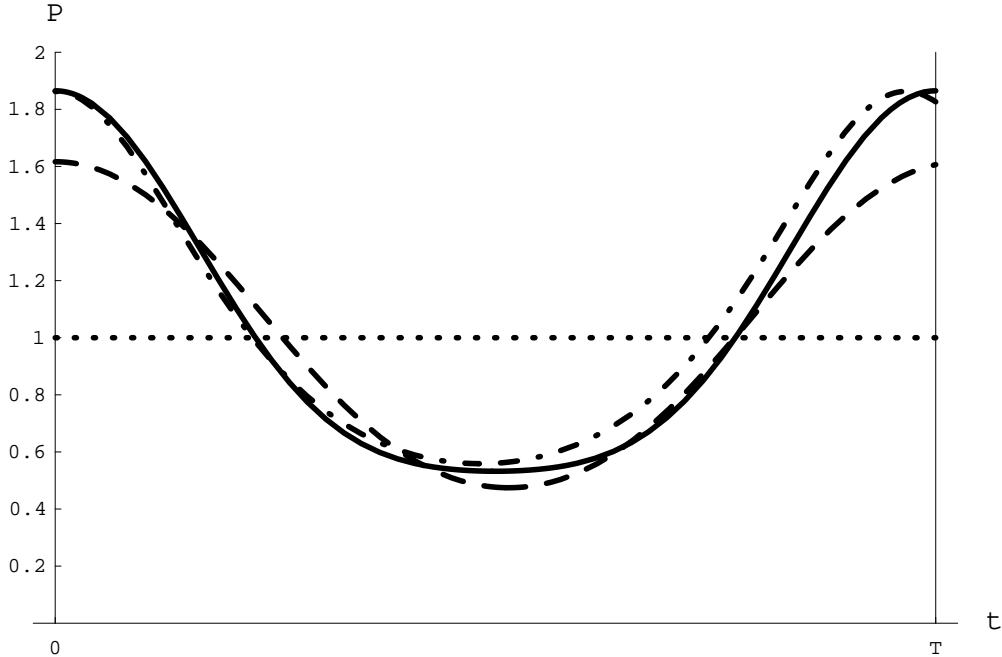}
\end{center}
\caption{{\protect\footnotesize The total power $P$ in four cases as
function of $t$ during one orbital period $T$, with $M=m_{1}$. The dotted
line is the circular orbit, i.e., $e=n_{0}^{r}=0$ case, and its total power $%
P$ is used as a reference to the others. With $e=0.10$ and $\frac{m_{1}}{a}
=0.02$, the total power $P$ for elliptic Keplerian orbit is represented by
the dot-dashed line. The dashed line is $P$ with $x^{\protect\mu }$ with
first-order geodesic deviation, and $\frac{n_{0}^{r}}{R}=0.10$ and $\frac{M}{
R}=0.02$. Finally, the total power $P $ with second-order geodesic deviation
is represented by the solid line, where $\frac{n_{0}^{r}}{R}=0.10$ and $%
\frac{M}{R}=0.0201$.}}
\end{figure}

Because the emitted total powers $P$ calculated with geodesic deviations
depend on the $\frac{M}{R}$ ratio, we see that the period is not $T=\frac{%
2\pi a^{3/2}}{\sqrt{Gm_{1}}}$ (third Kepler's law), but an increased one, 
\begin{equation}
T=\frac{2\pi R^{3/2}}{\sqrt{GM}\sqrt{1-\frac{6GM}{R}}}+{\cal O}\left( \frac{%
(n_{0}^{r})^{2}}{R^{2}}\right) .  \label{Trel}
\end{equation}
This effect is the direct consequence of the form of angular frequency $%
\omega $ that appears in the first and higher-order geodesic deviations. 
\newline
\indent
Another expected feature of Figures $1$ and $2$ : as $e$ (i.e., $\frac{%
n_{0}^{r}}{R}$) is kept small, the $P$ using geodesic deviations converge
very fast in respect of the orders of geodesic deviation. \newline
\indent
Caution is required as the use of quadrupole approximation is not allowed
for high values of $\frac{M}{R}$, so the exact amplitude and shape of $P$
using geodesic deviations can only be calculated if additional $\frac{M}{R}$
contributions to the gravitational radiation formula are included. This
approach, but using the post-Newtonian expansion scheme, is well developed
in Refs. \cite{Tanaka1,Tanaka2,Poisson}.

\section{Discussion}

In this paper we have introduced an extension of the geodesic deviation idea
in order to calculate approximate orbits of point masses in gravitational
fields. This scheme is of practical applicability to the problem of the
emission of gravitational radiation. Although in the present paper we
restricted our investigations to the case of Schwarzschild background
fields, our method can be easily extended to other background fields \cite
{KMMH}. An example is provided by the discussion of Reissner-Nordstr{\o}m
fields in Ref.\ \cite{BHK}. \newline
\indent
Since the initial work by Einstein, the problem of orbits and radiation is
addressed in the literature mostly through the post-Newtonian expansion
scheme \cite{Peters}, \cite{Damourruelle}-\cite{Steinbauer}. In this
approach the starting point for the successive approximations is found in
Newtonian theory, whilst relativistic effects are introduced by corrections
of higher order in $\frac{v}{c}$ or $\frac{M}{R}$. The advantage of this
approach is, that one can start with an orbit of arbitrary high
eccentricity. In contrast, our approach starts from a true solution of the
relativistic problem, but in the case of the Schwarzschild background we
choose a circular one. We then approach finite eccentricity orbits in a
fully relativistic scheme, by summing up higher-order geodesic deviations,
for which we have derived the explicit expressions.

The two approaches are complementary in the following sense: the
post-Newt\-onian scheme gives better results for small values of $\frac{M}{R}
$ and arbitrary eccentricity, whereas our scheme is best adapted for small
eccentricities, but arbitrary values of $\frac{M}{R}<\frac{1}{6}$. In both
approaches the emission of gravitational radiation is estimated using the
quadrupole formula, based on a flat-space approximation.

The next challenge is to include finite-size and radiation back-reaction
effects. In the post-Newtonian scheme some progress in this direction has
already been made. In this aspect our result may be regarded as the first
term in an expansion in $\frac{m}{M}$. Other applications can be found in
problems of gravitational lensing and perturbations by gravitational waves.

\vskip0.4cm \noindent {\tbf Acknowledgments}\newline
\newline
\noindent
R. K.\ and R. C. Jr.\ wish to thank Dr.\ Christian Klein for
enlightening discussions and C\'{e}dric Leygnac for his valuable help in
checking the results. We also wish to express our thanks to the Referees
for their constructive criticisms and useful remarks that helped to improve
the presentation.
\newpage

\section*{Appendix 1}

The covariant third-order deviation equation is obtained via the same
procedure that has served to derive the second-order covariant deviation.
The third-order geodesic deviation itself is 
\begin{eqnarray}
\delta ^{3}x^{\mu } &=&{(p-p_{0})^{3}\,\frac{\partial ^{3}x^{\mu }}{\partial
p^{3}}}  \nonumber \\
&=&\epsilon ^{3}{\left[ h^{\mu }-3\Gamma _{\lambda \nu }^{\mu }n^{\lambda
}b^{\nu }+\left( \partial _{\kappa }\Gamma _{\lambda \nu }^{\mu }-2\Gamma
_{\lambda \sigma }^{\mu }\Gamma _{\kappa \nu }^{\sigma }\right) n^{\kappa
}n^{\lambda }n^{\nu }\right] ,}  \label{a1.1}
\end{eqnarray}
where $h^{\mu }=Db^{\mu }/Dp=D^{2}n^{\mu }/Dp^{2}$. We derive the
third-order deviation equation by taking the covariant derivative w.r.t.\ $p$
of Eq.\ (\ref{geodev2b}) for $b^{\mu }$, with the result 
\begin{equation}
\begin{array}{ll}
\displaystyle{\ \frac{D^{2}h^{\mu }}{Ds^{2}}} & +\,R_{\rho \lambda \sigma
}^{\;\;\;\;\;\mu }u^{\lambda }u^{\sigma }h^{\rho }\,=\,\displaystyle{\
6R_{\lambda \rho \sigma }^{\;\;\;\;\;\mu }(u^{\lambda }n^{\rho }\frac{%
Db^{\sigma }}{Ds}+u^{\lambda }\frac{Dn^{\sigma }}{Ds}b^{\rho })} \\ 
&  \\ 
& \displaystyle{\ ~+\nabla _{\tau }\nabla _{\nu }R_{\lambda \rho \sigma
}^{\;\;\;\;\;\mu }(u^{\nu }u^{\rho }n^{\tau }n^{\lambda }n^{\sigma }+u^{\rho
}u^{\sigma }n^{\tau }n^{\nu }n^{\lambda })} \\ 
&  \\ 
& \displaystyle{\ ~+\nabla _{\nu }R_{\lambda \rho \sigma }^{\;\;\;\;\;\mu }%
\left[ 4u^{\lambda }n^{\nu }n^{\rho }\frac{Dn^{\sigma }}{Ds}+u^{\nu }u^{\rho
}n^{\lambda }b^{\sigma }+u^{\rho }u^{\sigma }n^{\nu }b^{\lambda }\right. }
\\ 
&  \\ 
& \displaystyle{\ ~\left. +2\left( u^{\rho }n^{\lambda }n^{\nu }\frac{%
Dn^{\sigma }}{Ds}+u^{\nu }n^{\lambda }n^{\sigma }\frac{Dn^{\rho }}{Ds}%
+u^{\nu }u^{\rho }n^{\lambda }\frac{Db^{\sigma }}{Ds}+u^{\rho }u^{\sigma
}n^{\nu }\frac{Db^{\lambda }}{Ds}\right) \right] } \\ 
&  \\ 
& \displaystyle{\ ~+4R_{\lambda \rho \sigma }^{\;\;\;\;\;\mu }n^{\rho }\frac{%
Dn^{\lambda }}{Ds}\frac{Dn^{\sigma }}{Ds}+4R_{\lambda \rho \sigma
}^{\;\;\;\;\;\mu }R_{\alpha \beta \gamma }^{\;\;\;\;\;\sigma }u^{\lambda
}u^{\beta }n^{\gamma }n^{\alpha }n^{\rho }.}
\end{array}
\label{geodev3b}
\end{equation}
Related studies of higher-order differentials and their covariant
generalizations from a more general perspective can be found in recent
papers \cite{Kerner98,Abramov}.

\section*{Appendix 2}

\indent
The contributions of various orders to the geodesic deviation obtained
in Section $2$ can be deduced in an elegant, coordinate-independent and
slightly more general manner \cite{Kobayashi}. Given a one-parameter
congruence of geodesics, one can define the tangent vector field $Z$ and
the local Jacobi field $X$; then the Lie bracket of these fields
vanishes (because the congruence spans a submanifold and therefore is
integrable), so that $[X,Z]=0$.\newline
\indent
The geodesic equation is $\nabla _{Z}\,Z=0$. Then, applying the definition
of Riemann tensor to the vectors $X,Y$ and $Z$: 
\begin{equation}
\left[ \nabla _{X}\nabla _{Z}-\nabla _{Z}\nabla _{X}\right] \,Z-\nabla
_{\lbrack X,Z]}\,Z=R(X,Z)\,Z,
\end{equation}
and taking into account that $[X,Z]=0$ as well as the fact that $\nabla
_{X}Z=\nabla _{Z}X$ and the anti-symmetry of $R(X,Z)$ in its two arguments,
we get easily 
\begin{equation}
\nabla _{Z}\nabla _{X}\,Z=\nabla _{Z}^{2}\,X=R\,(Z,X)\,Z,  \label{elegant2}
\end{equation}
which coincides with Eq.\ (\ref{geodev1a}) for the geodesic deviation, after
we identify the components of the vector fields as $Z^{\mu }=u^{\mu},\,
X^{\mu}= n^{\mu }$.

One may continue in the same spirit and introduce two linearly independent
Jacobi fields, $X$ and $Y$, both satisfying $[X,Z]=0=[Y,Z]$, to obtain the
coordinate-independent form of Eq.\ (\ref{geodev2b}), as follows. The two
linearly independent Jacobi fields, $X$ and $Y$, satisfy $[X,Z]=0$ and $%
[Y,Z]=0$. By virtue of the Jacobi identity, we have $[[X,Y],Z]=0$, hence $%
[X,Y]$ is also a Jacobi field (i.e., it satisfies Eq.\ (\ref{elegant2})).
Applying the same formula to this field, we get 
\begin{equation}
\nabla _{Z}^{2}\,\left( [X,Y]\right) =R(Z,[X,Y])\,Z  \label{elegant3}
\end{equation}
Then, using the fact that $\nabla _{X}\,Y-\nabla _{Y}\,X=[X,Y]$, we can
write the left hand side of the above equation as $\nabla _{Z}^{2}\left(
\nabla _{X}\,Y-\nabla _{Y}\,X]\right)$.

Next, we can write this equation explicitly as 
\[
\nabla^2_Z \, (\nabla_X Y - \nabla_Y X) = R(Z, \, \nabla_XY - \nabla_YX) \,
Z 
\]
and furthermore, using the linearity property, as 
\begin{equation}
\nabla^2_Z (\nabla_Y X) - R(Z, \nabla_Y X) \, Z = \nabla^2_Z (\nabla_X Y) -
R(Z, \nabla_X Y) \, Z.
\end{equation}
The left-hand side of the above equation coincides with the usual Jacobi
equation applied to the field $\nabla_Y X$, whereas the right-hand side can
be transformed using the definition of the Riemann tensor: the term $%
\nabla^2_Z (\nabla_X Y)$ gives 
\begin{equation}
\begin{array}{lll}
\nabla^2_Z Y & = & \nabla_Z\, (\nabla_Z \nabla_X Y) = \nabla_Z (\nabla_Z
\nabla_X Y - \nabla_X \nabla_Z Y) + \nabla_Z(\nabla_X \nabla_Z Y) \\ 
&  &  \\ 
& = & \nabla_Z\, (R (Z,X)\, Y) + \nabla_Z(\nabla_X \nabla_Z Y) \\ 
&  &  \\ 
& = & \left[ (\nabla_Z R)(Z,X) \right] Y + R(Z,\nabla_Z X) Y + R(Z,X)\,
\nabla_Z Y + \nabla_Z(\nabla_X \nabla_Z Y)
\end{array}
\end{equation}
(here we used the fact that $\nabla_Z Z = 0$). Manipulating further in the
same manner the commutators of covariant derivations, we arrive at the
result 
\[
\nabla _{Z}^{2}\,(\nabla _{Y}\,X)-R(Z,\nabla _{Y}\,X)\,Z= 
\]
\begin{equation}
\nabla _{X}\,R(Z,Y)\,Z+\nabla _{Z}\,R(Z,X)\,Y+2R(Z,Y)\nabla
_{Z}\,X+2R(Z,X)\,\nabla _{Z}\,Y
\end{equation}
which is equivalent to Eq.\ (\ref{geodev2b}) upon identification $Y=X$ and $%
b^{\mu} = (\nabla_X X)^{\mu}$. 

Although these coordinate-independent derivations are more elegant,
their results are not so useful for pratical computations, i.e., the
non-manisfestly covariant form of the results is better adapted for the
calculus of successive deviations in a given local coordinate system.

\section*{Appendix 3}

{\it Connections and curvatures for Schwarzschild geometry}\newline
In this appendix we collect the expressions for the components of the
connections and Riemann curvature used in the main body of the paper. 
\vskip
0.2cm \noindent {\it A.\ Connections.} From the line-element (\ref{Schmetric})
one derives the following expressions for the connection coefficients: 
\begin{equation}
\Gamma _{\rho \lambda }^{\mu }=\frac{1}{2}\,g^{\mu \sigma }\,(\partial
_{\rho }g_{\sigma \lambda }+\partial _{\lambda }g_{\rho \sigma }-\partial
_{\sigma }g_{\rho \lambda });  \label{Christoffel}
\end{equation}
\begin{equation}
\begin{array}{lll}
\Gamma _{rt}^{t} & = & \displaystyle{-\,\Gamma _{rr}^{r}=\frac{M}{r^{2}{%
\left( 1-\frac{2M}{r}\right) }},{\hspace{5.9em}}\Gamma _{tt}^{r}\,=\,\frac{M%
}{r^{2}}\,\left( 1-\frac{2M}{r}\right) }, \\ 
&  &  \\ 
\Gamma _{\varphi \varphi }^{r} & = & \displaystyle{-\,r\,\sin ^{2}\,\theta
\,\left( 1-\frac{2M}{r}\right) ,\hspace{4.5em}\Gamma _{\theta \theta
}^{r}\,=\,-\,r\,\left( 1-\frac{2M}{r}\right) }, \\ 
&  &  \\ 
\Gamma _{r\theta }^{\theta } & = & {\displaystyle\Gamma _{r\varphi
}^{\varphi }\,=\,\frac{1}{r},\hspace{3em}\Gamma _{\theta \varphi }^{\varphi
}\,=\,\frac{\cos \theta }{\sin \theta },\hspace{3em}\Gamma _{\varphi \varphi
}^{\theta }\,=\,-\sin \theta \,\cos \theta .}
\end{array}
\label{a.1}
\end{equation}
\noindent {\it B.\ Curvature components.} The corresponding curvature
two-form components $R_{\mu \nu }=\frac{1}{2}\,R_{\kappa \lambda \mu \nu
}dx^{\kappa }\wedge dx^{\lambda }$ are: 
\begin{equation}
\begin{array}{lll}
R_{tr} & = & {\displaystyle\frac{2M}{r^{3}}\,dt\wedge dr,\hspace{8.8em}}%
R_{t\theta }{\,=\,-\frac{M}{r}\,\left( 1-\frac{2M}{r}\right) \,dt\wedge
d\theta ,} \\ 
&  &  \\ 
R_{t\varphi } & = & \displaystyle{-\frac{M}{r}\,\left( 1-\frac{2M}{r}\right)
\,\sin ^{2}\theta \,dt\wedge d\varphi ,\hspace{1.7em}}R_{r\theta }{\,=\,%
\frac{M}{r{\left( 1-\frac{2M}{r}\right) }}\,dr\wedge d\theta ,} \\ 
&  &  \\ 
R_{r\varphi } & = & {\displaystyle\frac{M}{r(1-\frac{2M}{r})}\,\sin
^{2}\theta \,dr\wedge d\varphi ,\hspace{2.2em}}R_{\theta \varphi }{\
\,=\,-\,(2Mr)\,\sin ^{2}\,\theta \,d\theta \wedge d\varphi .}
\end{array}
\label{a3.2}
\end{equation}

\end{document}